\documentstyle[12pt]{article}

\newlength{\dinwidth}
\newlength{\dinmargin}
\begin{document}
\titlepage
\begin{flushright}
DTP/96/26  \\
March 1996 \\
\end{flushright}

\begin{center}
\vspace*{2cm}
{\Large \bf Deep inelastic events containing a forward photon as
a probe of small $x$ dynamics} \\
\vspace*{1cm}
J.\ Kwieci\'{n}ski\footnote{On leave from Henryk
Niewodnicza\'{n}ski Institute of Nuclear Physics, 31-342
Krak\'{o}w, Poland.}, S.\ C.\ Lang and A.\ D.\ Martin, \\
\vspace*{0.5cm}
Department of Physics, University of Durham, Durham, DH1 3LE,
England
\end{center}

\vspace*{4cm}
\begin{abstract}
We calculate the rate of producing deep inelastic events
containing an energetic isolated forward photon at HERA.  We
quantify the enhancement arising from the leading $\log 1/x$
gluon emissions with a view to using such events to identify the
underlying dynamics.
\end{abstract}

\newpage
\noindent {\large \bf 1.  Introduction} \\

In the deep-inelastic small (Bjorken) $x$ regime the gluon is the
dominant partonic constituent of the proton, and the behaviour of
the observables can be predicted in terms of the gluon
distribution.  At sufficiently small $x$ it is necessary to resum
the $\log 1/x$ contributions which, to leading-order accuracy, is
accomplished by the BFKL equation \cite{bfkl} for the gluon
distribution, $f (x, k_T^2)$, unintegrated over the gluon
transverse momentum $k_T$.  As $x$ decreases the solution $f (x,
k_T^2)$ is characterized by a singular $x^{-
\lambda}$ growth accompanied by a diffusion in $\ln k_T^2$.  To
leading $\log 1/x$ accuracy, the \lq\lq BFKL intercept" is
$\lambda
= (3 \alpha_S/\pi) \: 4 \ln 2$ for fixed $\alpha_S$ (or $\lambda
\approx 0.5$ if a reasonable prescription is used to allow
$\alpha_S$ to run in the BFKL equation \cite{akms}).  Sub-leading
effects are expected to reduce the exponent.

The singular $x^{- \lambda}$ behaviour feeds through, via the
$k_T$-factorization theorem, into the proton structure function
$F_2 (x, Q^2)$.  Thus, in principle, measurements of $F_2$ at
HERA should be able to determine $\lambda$.  However, in practice
it is difficult to distinguish the BFKL behaviour from that
predicted by conventional DGLAP evolution, which resums $\alpha_S
\log Q^2$ contributions.  A major part of the problem is the
necessity to provide non-perturbative input.  Over the HERA
regime, DGLAP evolution can mimic the steep $x^{- \lambda}$
growth
by either taking a valence-like gluon $xg (x, Q_0^2) \rightarrow
0$ as $x \rightarrow 0$ at some sufficiently low input scale
$Q_0^2$ \cite{grv}, or by assuming an input form $xg (x, Q_0^2)
\sim x^{- \lambda}$ as $x \rightarrow 0$ \cite{mrs,cteq}.  Here
$g (x, Q^2)$ is the conventional gluon distribution (integrated
over $k_T$), namely
\begin{equation}
g (x, Q^2) \; = \; \int^{Q^2} \: \frac{dk_T^2}{k_T^2} \; f (x,
k_T^2).
\label{eq:a1}
\end{equation}
On the other hand, in the BFKL description of $F_2$, there is a
diffusion in $\ln k_T^2$ accompanying the translation of the
input
gluon $f (x, k_T^2)$ to smaller $x$ values.  This leads to a
significant contribution from the infrared region which is beyond
the scope of perturbative QCD and which has to be included
phenomenologically.

Mueller \cite{m} proposed an elegant way to overcome these
ambiguities, and so sharpen up the theoretical description.  He
suggested that the study of deep-inelastic $(x,
Q^2)$ events which contain an identified energetic forward jet
$(x_j, k_{jT}^2)$ would, in principle, allow the identification
of the $x^{- \lambda}$ behaviour, or to be precise the
$(x/x_j)^{- \lambda}$ behaviour as $x/x_j \rightarrow 0$.  The
process is illustrated in Fig.\ 1(a)\footnote{Other properties of
the final state have been proposed as possible ways to expose the
mutual relationship between the steep power-like increase of the
cross section and the lack of ordering in the transverse momenta
of the gluon emissions along the chain.  These include studies of
(i) the transverse energy flow \cite{gkms}, (ii) the azimuthal
decorrelation of two jets separated by a large rapidity gap
\cite{dd,wjs}, and (iii) the azimuthal decorrelation of electro-
or photo-produced dijets \cite{fr,a}.  The first tends to be
masked by uncertainties due to hadronization and the other two by
decorrelations which arise from other sources than $\log (1/x)$
resummation.}.  Ideally we would select
deep-inelastic events containing a jet whose transverse momentum
$k_{jT}$ is sufficiently large to suppress the diffusion into the
infrared region, and with longitudinal momentum fraction $x_j$ as
large as experimentally feasible so that $x/x_j$ is as small as
possible, typically $x_j > 0.05$.  According to BFKL dynamics the
differential structure function for this process has leading
small $x/x_j$
behaviour
\begin{equation}
\frac{\partial F_2}{\partial (\log 1/x_j) \partial k_{jT}^2} \; =
\; C \: \alpha_S (k_{jT}^2) \: x_j \left [g + \frac{4}{9} \: (q +
\overline{q}) \right ] \; \left ( \frac{x}{x_j} \right )^{-
\lambda}
\label{eq:a2}
\end{equation}
where the parton distributions ($g, q$ and $\overline{q}$) are to
be evaluated at $(x_j, k_{jT}^2)$ --- where they are well known. 
The normalisation coefficient $C$ is given in refs.\
\cite{t,bd,kms1,kms2}. 
Basically the idea is to try to identify the BFKL behaviour from
deep-inelastic scattering off a known {\it parton} at
large $k_{jT}^2$, as opposed to scattering off a
{\it proton} for which we need to specify its non-perturbative
structure at small $k_T^2$.  In practice, however,
the clean identification and kinematic measurement of a forward
jet, closely adjacent to the proton
remnants, poses a severe experimental challenge, particularly for
jets at the smaller values of $k_{jT}^2$ where the DIS + jet
events are more
numerous.  Nevertheless experimental studies \cite{h1} have been
attempted with encouraging results, albeit for rather low values
of $k_{jT}^2$.

Here we study a related, but alternative process --- namely
deep-inelastic events containing an identified forward photon,
see
Fig.\ 1(b).  As a probe of small $x$ dynamics, the process DIS +
$\gamma$ has both advantages and
disadvantages as compared to DIS + jet.  A major advantage is
that the measurement of a photon should be cleaner than that of a
jet, and less ambiguous, particularly at the lower values of
$k_T$. 
Moreover since the $q$ or $\overline{q}$ jet (denoted
$x_q^\prime$ in
Fig.\ 1(b)) is not identified we can enlarge the data sample by
including events in which its constituents mingle with the proton
remnants. 
On the other hand we expect the DIS + $\gamma$ rate
to be suppressed by a factor of order $\alpha/2\pi$ relative to
DIS + jet, though we should gain a little back since the photon
is easier to measure than the jet and the recoiling quark jet is
not identified.  A second disadvantage is that we require
isolation of the photon so as to avoid events where it arises
from the decay of a parent $\pi^0$ etc. 

We first present (in Section 2) the QCD formalism necessary to
calculate the cross section for the DIS + forward photon process
at small $x$.  Then in Section 3 we discuss the acceptance
criteria to be applied to the outgoing photon.  In Section 4 we
quantify the DIS + $\gamma$ event rate expected in the
experiments at the HERA electron-proton collider.  Finally
Section 5 contains a brief discussion. \\

\bigskip
\noindent {\large \bf 2.  Basic QCD formula for the DIS +
$\gamma$ process} \\

Here we study prompt photon production in deep-inelastic
scattering at small $x$.  That is the process
\begin{equation}
\lq\lq \gamma" \: + \: p \; \rightarrow \; \gamma (x_\gamma,
k_{\gamma T}) \: + \: X,
\label{eq:a3}
\end{equation}
sketched in Fig.\ 1(b), in which the photon is identified in the
final state.  Deep inelastic events with small $x$ and large
$x_\gamma$ offer an opportunity to identify the effects of the
BFKL resummation of the $\alpha_S \ln (x_\gamma/x)$
contributions, which arise from the sum over the real and virtual
gluon emissions, such as the one depicted in Fig.\ 1(b).  In
analogy to the DIS + jet process, the advantage of process
(\ref{eq:a3}) is that the outgoing photon acts as a trigger to
select events in which the deep-inelastic scattering occurs off a
quark in a kinematic region where its distribution, $q (x_q,
k_{\gamma T}^2)$, is known.

The differential structure functions for this process may be
written in the form
\begin{equation}
\frac{\partial^2 F_i}{\partial x_\gamma \partial k_{\gamma T}^2}
\; = \; \int \: \frac{d^2 k_{gT}}{\pi k_{gT}^4} \: \int \:
\frac{dx_q}{x_q} \: \Phi_i \left (\frac{x}{x_q}, k_{gT}^2, Q^2
\right ) \sum_q e_q^2 \: \left [ q (x_q, k_{\gamma T}^2) \: + \:
\overline{q} (x_q, k_{\gamma T}^2) \right ] \; \frac{|{\cal
M}|^2}{z_\gamma z_q^\prime}
\label{eq:a4}
\end{equation}
for $i = T, L$.  The variables are indicated on Fig.\ 1(b) and
are defined below.  The region of integration in (\ref{eq:a4})
will be specified later.  The subprocess $q (k_q) + g (k_g)
\rightarrow
q (k_q^\prime) + \gamma (k_\gamma)$ (and also $\overline{q}g
\rightarrow \overline{q}\gamma$) is described by the two Feynman
diagrams shown in Fig.\ 2.  It has matrix element squared
\begin{equation}
|{\cal M}|^2 \; = \; 2 \: C_2 (F) \; \frac{\alpha_S (k_{\gamma
T}^2)}{2
\pi} \; \frac{\alpha}{2 \pi} \; \frac{k_{gT}^2 [1 \: + \: (1 -
z_\gamma)^2]}{\hat{s} (- \hat{u})}
\label{eq:a5}
\end{equation}
where $C_2 (F) = 4/3$ and the invariants $\hat{s}$ and $\hat{u}$
are
\begin{equation}
\hat{s} \; = \; (k_\gamma \: + \: k_q^\prime)^2, \;\;\;\;\;
\hat{u} \; = \; (k_q \: - \: k_\gamma)^2.
\label{eq:a6}
\end{equation}
The fractional momenta
$x_i$ in (\ref{eq:a4}) are defined by the Sudakov decomposition
of the particle 4-momenta
\begin{equation}
k_i \; = \; x_i  p^\prime \: + \: \beta_i q^\prime \: + \:
\mbox{\boldmath $k$}_{iT}
\label{eq:a7}
\end{equation}
where at high energies $p^\prime$ and $q^\prime$ are to a good
approximation the light-like 4-momenta
associated with the incoming proton and virtual photon
respectively
\begin{equation}
p^\prime \; \equiv \; p \: - \: \frac{M^2}{2p.q} \: q, \;\;\;\;
q^\prime \; \equiv \; q \: + \: xp,
\label{eq:a8}
\end{equation}
where $M$ is the mass of the proton.  The variables $z_i$ in
(\ref{eq:a4}) and (\ref{eq:a5}) are given by
\begin{equation}
z_i \; = \; x_i/x_q.
\label{eq:a9}
\end{equation}
The 4-momenta of the outgoing photon and quark jet satisfy the
on-mass-shell condition $k_i^2 = 0$ which gives
\begin{equation}
\beta_i \; = \; \frac{x}{x_i} \: \frac{k_{iT}^2}{Q^2}
\label{eq:b9}
\end{equation}
for $i = \gamma$ or the outgoing quark $q^\prime$.

The functions $\Phi_i (z, k^2, Q^2)$ in (\ref{eq:a4}) describe
the gluon chain which couples to the incoming virtual photon. 
The function $\Phi_T$ corresponds to the function $F$ in ref.\
\cite{kms1}.  To be precise, the factors $\Phi_i/k^2$ can be
identified with the virtual gluon structure functions integrated
over the longitudinal momentum of the gluon.  They can be
obtained from the BFKL equation with the boundary condition given
by the quark box (and \lq\lq crossed" box) contributions.  That
is
\begin{eqnarray}
\Phi_i (z, k^2, Q^2) & = & \Phi_i^{(0)} (z, k^2, Q^2) \: +
\nonumber \\
& + & \frac{3 \alpha_S}{\pi} \: k^2 \: \int_z^1 \:
\frac{dz^\prime}{z^\prime} \: \int_0^\infty \:
\frac{dk^{\prime 2}}{k^{\prime 2}} \; \left [ \frac{\Phi_i
(z^\prime, k^{\prime 2}, Q^2) \: - \: \Phi_i (z^\prime, k^2,
Q^2)}{| k^{\prime 2} \: - \: k^2 |} \: + \: \frac{\Phi_i
(z^\prime, k^2, Q^2)}{\sqrt{4k^{\prime 4} \: + \: k^4}} \right ]
\nonumber \\
& & 
\label{eq:a10}
\end{eqnarray}
where the inhomogeneous or driving terms $\Phi_i^{(0)}$
correspond to the sum of the quark box and crossed-box
contributions.  At small $z$ we have
\begin{equation}
\Phi_i^{(0)} (z, k^2, Q^2) \; \approx \; \Phi_i^{(0)} (z = 0,
k^2, Q^2) \; \equiv \; \Phi_i^{(0)} (k^2, Q^2)
\label{eq:a11}
\end{equation}
where
\begin{eqnarray}
\Phi_T^{(0)} (k^2, Q^2) & = & 2 \: \sum_q \: e_q^2 \:
\frac{Q^2}{4 \pi^2} \: \alpha_S \: \int_0^1 \: d \beta \: \int \:
d^2 \kappa \left [\beta^2 \: + \: (1 - \beta)^2 \right ] \left (
\frac{\kappa^2}{D_1^2} \: - \: \frac{\mbox{\boldmath $\kappa$} .
(\mbox{\boldmath $\kappa$} - \mbox{\boldmath $k$})}{D_1 D_2}
\right ) \nonumber \\
& & \nonumber \\
\Phi_L^{(0)} (k^2, Q^2) & = & 2 \sum_q \: e_q^2 \;
\frac{Q^4}{\pi^2} \: \alpha_S \: \int_0^1 \: d \beta \: \int \:
d^2 \kappa \: \beta^2 (1 - \beta)^2 \; \left ( \frac{1}{D_1^2} \:
- \: \frac{1}{D_1 D_2} \right )
\label{eq:12}
\end{eqnarray}
with the denominators $D_i$ given by
\begin{eqnarray}
D_1 & = & \kappa^2 \: + \: \beta (1 - \beta) \: Q^2 \nonumber \\
& & \\
D_2 & = & (\mbox{\boldmath $\kappa$} - \mbox{\boldmath $k$})^2 \:
+ \: \beta (1 - \beta) \: Q^2. \nonumber
\label{eq:a13}
\end{eqnarray}

The BFKL equation, (\ref{eq:a10}), has been written for fixed QCD
coupling $\alpha_S$.  In this case the leading small $x$
behaviour of the solution can be given analytically.  We have
\begin{eqnarray}
\Phi_T (z, k^2, Q^2) & = & \frac{9 \pi^2}{512} \: \frac{2 \sum
e_q^2 \alpha_S^{\frac{1}{2}}}{\sqrt{21 \zeta (3)/2}} \; (k^2
Q^2)^{\frac{1}{2}} \; \frac{z^{- \alpha_P + 1}}{\sqrt{\ln (1/z)}}
\; \left [ 1 \: + \: {\cal O} \: \left ( \frac{1}{\ln (1/z)}
\right ) \right ] \nonumber \\
& & \\
\Phi_L (z, k^2, Q^2) & = & \frac{2}{9} \: \Phi_T (z, k^2, Q^2)
\nonumber
\label{eq:b13}
\end{eqnarray}
where we have omitted the Gaussian diffusion factor in $\ln
(k^2/Q^2)$.  The Riemann zeta function $\zeta (3) = 1.202$, and
$\alpha_P$, the celebrated BFKL intercept, is given by
\begin{equation}
\alpha_P \: - \: 1 \; = \; \frac{12 \alpha_S}{\pi} \; \ln 2.
\label{c13}
\end{equation}

In practice we allow the coupling $\alpha_S$ to run, following
the prescription of ref.\ \cite{kms1}.  That is we solve the BFKL
equations for
\begin{equation}
H_i (z, k^2, Q^2) \; = \; \frac{3 \alpha_S (k^2)}{\pi} \; \Phi_i
(z, k^2, Q^2).
\label{eq:a14}
\end{equation}
Actually we solve the differential form of the equations
\begin{equation}
\frac{\partial H_i (z, k^2, Q^2)}{\partial \log (1/z)} \; = \;
\frac{3 \alpha_S (k^2)}{\pi} \: k^2 \: \int_{k_0^2}^\infty \:
\frac{dk^{\prime 2}}{k^{\prime 2}} \; \left [ \frac{H_i (z,
k^{\prime 2}, Q^2) - H_i (z, k^2, Q^2)}{| k^{\prime 2} - k^2 |}
\: + \: \frac{H_i (z, k^2, Q^2)}{\sqrt{4k^{\prime 4} \: + \:
k^4}} \right ]
\label{eq:a15}
\end{equation}
subject to the boundary conditions
\begin{equation}
H_i (z = z_0, k^2, Q^2) \; = \; H_i^{(0)} (k^2, Q^2)
\label{eq:a16}
\end{equation}
where we choose $z_0 = 0.1$ and the cut-off $k_0^2 = 1$ GeV$^2$. 
For any small $z$ the solutions $H_i (z, k^2, Q^2)$ therefore
only depend on the behaviour of $H_i$ in the interval $(z, z_0)$.
We impose the same cut, $x/x_q < 0.1$, in the $x_q$ integration
in (\ref{eq:a4}).  Of course the integration is also subject to
the kinematic constraint $x_q > x_\gamma$.  We find that the
slope
$\lambda = \alpha_P - 1$ and the overall normalization of the
solution of the BFKL equation with running coupling
$\alpha_S$ are in general smaller than those obtained with fixed
coupling $\alpha_S (Q^2)$.

We are now in a position to calculate the differential structure
function $\partial F_2/\partial x_\gamma \partial k_{\gamma T}^2$
from (\ref{eq:a4}).  We compute it first using $\Phi_2$
obtained from solving the BFKL equation (\ref{eq:a15}) and then
repeat the calculation using for $\Phi_2$ just the driving (quark
box) term $\Phi_2^{(0)}$.  The difference between the two
calculations
is a measure of the effect of the BFKL resummation of soft gluon
emissions.  The result for the $F_2$ differential structure
function is shown in Fig.\ 3 as a function of $x_\gamma$ for
three different values\footnote{The result using $\Phi_2 =
\Phi_2^{(0)}$ is independent of $x$ (until we reach the kinematic
limit $\beta_\gamma = 1$), and its shape in $x_\gamma$
reflects the quark and antiquark distributions in the proton
after integration over $x_q$.  In contrast for DIS + jet the
$x_j$ shape of the box contribution is dominated by the gluon
distribution $x_j g (x_j, k_{jT}^2)$, that is the gluon is
sampled \lq\lq locally".} of $x$.  We see that the difference
between the BFKL and quark box results is dramatic once $x$ is
sufficiently small $(x {\lower .7ex\hbox{$\;\stackrel{\textstyle 
<}{\sim}\;$}} 10^{-3})$ and provided
$x_\gamma \sim 0.1$.  Indeed we require $x/x_\gamma$ to be small
for the formalism to be valid.  As explained in ref.\ \cite{kms1}
it is the shape rather than the magnitude of $\partial
F_i/\partial x_\gamma \partial k_{\gamma T}^2$ which is the more
reliable discriminator of the underlying dynamics at small $x$,
since the normalization of the QCD predictions is subject to
uncertainties arising mainly from the regions of low transverse
momenta.

Fig.\ 3 for the differential structure function for DIS +
$\gamma$ events
should be compared with a similar figure for DIS + jet events
shown in Fig.\ 7 of ref.\ \cite{kms1}.  We see that the cross
section is about a factor of 1000 lower for the photon process,
as could be anticipated from the presence of the extra $\alpha/2
\pi$ coupling of the photon.  In the next section we will impose
reasonable experimental cuts on the outgoing photon, and then we
will quantify the event rate which may be observed at HERA. \\

\bigskip
\noindent {\large \bf 3.  Cuts to select the DIS + photon events}
\\

We are interested in semi-inclusive deep inelastic events in
which only the photon is measured in the final state, besides the
scattered electron.  On the other hand the integrals defining the
differential structure functions, (\ref{eq:a4}), contain a
potential singularity at $\hat{s} = 0$ which has to be
regularized by a suitably chosen cut-off.  A straightforward
limitation of the region of integration to $\hat{s} > s_0$, where
$s_0$ is some arbitrarily chosen value, is therefore not possible
for making a direct comparison of (\ref{eq:a4}) with experiment,
since $\hat{s}$ is only determined if the recoil quark jet
$q^\prime$ were to be detected as well as the photon.  However,
it is necessary to impose an isolation cut on the outgoing photon
to distinguish it from background photons which arise from the
decay of $\pi^0$'s produced either from the proton remnants or
the outgoing quark jet, $q^\prime$.  Suppose we impose an 
isolation
angle $\theta_0$ (where $\theta_0$ is chosen to be around, say,
$3^{\rm o}$--$10^{\rm o}$) defining a cone around the photon,
then
at the parton level we have $\theta_{\gamma q^\prime} >
\theta_0$ and hence a corresponding lower limit on $\hat{s}$,
since
\begin{equation}
\hat{s} \; > \; 2 E_\gamma E_{q^\prime} \: (1 - \cos
\theta_{\gamma q^\prime}).
\label{eq:a17}
\end{equation}

A second requirement is that our photon should be emitted in the
proton hemisphere in the virtual photon-proton c.m.\ frame to
avoid contamination from photons radiated from the
quark-antiquark pair which form the quark box.  Thus we require
\begin{equation}
x_\gamma \; > \; \beta_\gamma
\label{eq:a18}
\end{equation}
which on using (\ref{eq:b9}), gives
\begin{equation}
x_\gamma \; > \; \sqrt{x k_{\gamma T}^2/Q^2}.
\label{eq:a19}
\end{equation}
The hemisphere cut, when combined with the kinematic constraint
$x_q > x_\gamma$, imposes an implicit lower limit of the $x_q$
integration in (\ref{eq:a4})
\begin{equation}
x_q \; > \; x_\gamma \; > \; \sqrt{x k_{\gamma T}^2/Q^2}.
\label{eq:b19}
\end{equation}
We see that this limit is generally stronger than our BFKL
requirement that $x_q > 10 x$.

Finally there is the practical limitation that photons can only
be measured if they are emitted at sufficiently large angles to
the proton beam direction in the HERA laboratory frame, say
\begin{equation}
\theta_{\gamma p} \; > \; \overline{\theta}_0.
\label{eq:a20}
\end{equation}
The allowed regions of the kinematic variables $(x_\gamma,
k_{\gamma T}^2)$ describing the photon are shown in Fig.\ 3 for
various choices of $\overline{\theta}_0$.  We see that large
$x_\gamma$ photons are only emitted at small $\theta_{\gamma p}$;
for a given $\theta_{\gamma p}$ we can reach larger $x_\gamma$ by
observing photons with large $k_{\gamma T}^2$, but then the event
rate is depleted.  We also show on Fig.\ 3 the \lq\lq forward
hemisphere" boundary, (\ref{eq:a19}), for $x = 6 \times 10^{-4}$
and $Q^2 = 20$ GeV$^2$.

\newpage
\noindent {\large \bf 4.  Predictions for DIS + photon
production}

Here we present the numerical predictions for the cross section
for the production of an energetic photon in the deep inelastic
$ep$ scattering taking into account various cuts discussed in the
previous section.  The relevant cross section is given by the
following formula:
\begin{equation}
\frac{d \sigma}{dx_\gamma dk_{\gamma T}^2 dx dQ^2} \; = \;
\frac{4 \pi \alpha^2}{Q^4 x} \; \left [ (1 - y) \;
\frac{dF_2}{dx_\gamma dk_{\gamma T}^2} \; + \; \frac{y^2}{2} \:
\frac{dF_T}{dx_\gamma dk_{\gamma T}^2} \right ].
\label{eq:a21}
\end{equation}
We show in Fig.\ 5 the dependence of the integrated cross
section on the variation of (a) the angle $\theta_0$ defining the
photon isolation cone and (b) the $k_{\gamma T}^2$ threshold for
photon detection.  We see that the dependence on the isolation
cone angle is relatively weak.

In Fig.\ 6 we show the integrated cross section for prompt
forward photon
production as a function of $x$ for three different $Q^2$ bins: 
20-30, 30-40 and 40-50 GeV$^2$ respectively.  We compare the
predictions for the case where the BFKL small $x$ resummation is
incorporated with those where the gluon radiation is neglected. 
The $x$ dependence of the cross section is driven by the small
$z$ behaviour of the $\Phi_i$.  The results show a strong
enhanced increase with decreasing $x$ which is characteristic of
the effect of soft gluon resummation.  At $x \approx 10^{-4}$ the
cross section is about a factor of 3.5 larger than that in which
the BFKL effects are neglected.  Since the impact factors,
$\Phi_i^{(0)}$, are independent of $x$, the weak $x$ dependence
in the latter case arises mainly from acceptance.  Finally in
Fig.\ 7 we show the cross section in the various deep inelastic
$(x, Q^2)$ bins.  The values are shown in fb and so would
correspond to the number of events for an eventual integrated
luminosity of 1 fb$^{-1}$.

The DIS + $\gamma$ cross sections presented in Figs.\ 6 and 7 may
be compared directly with the values for the DIS + jet process
shown in Fig.\ 4 of \cite{kms2} and Fig.\ 8 of \cite{kms3}
respectively.  We see that there is a suppression of about a
factor of 400 in going from the DIS + jet to the DIS + $\gamma$
process.  Of course the ratio depends on the precise cuts that
are imposed in each case.  In table 1 we show the effect on the
DIS + $\gamma$ cross section in one of the $\Delta x, \Delta Q^2$
bins of varying $\overline{\theta}_0$ (the minimum angle to the
proton beam) and $\theta_0$ (the half-angle of the isolation
cone).  We see that the event rate is less sensitive to the cone
angle $\theta_0$ than to the minimum angle $\overline{\theta}_0$
to the beam.

\begin{table}[htb]
\caption{The DIS + $\gamma$ cross section in the bin $6 \times
10^{-4} < x < 8 \times 10^{-4}$, $20 < Q^2 < 30$ GeV$^2$ as
calculated in Fig.\ 7, but for different choices of
$\overline{\theta}_0$ and $\theta_0$.}
\begin{center}
\begin{tabular}{|c|c|c|} \hline
$\overline{\theta}_0$ & $\theta_0$ & $\Delta \sigma$ (fb) \\
\hline
$5^\circ$ & $3^\circ$ & 46.2 \\
$7^\circ$ & $3^\circ$ & 29.1 \\
$7^\circ$ & $5^\circ$ & 26.2 \\ \hline
\end{tabular}
\end{center}
\end{table}

\newpage
\noindent {\large \bf 5.  Discussion}

The DIS + jet process is, in principle, an ideal way to probe
small $x$ dynamics, {\it provided} sufficiently forward jets can
be measured.  In practice to separate cleanly such forward jets
from the proton remnants is a formidable challenge.  Here we have
studied the analogous DIS + $\gamma$ process which has the
advantage that forward photons can be more reliably measured than
forward jets, but for which the event rate is considerably
suppressed.  In section 4 we quantified the suppression.  There
we predicted the DIS + $\gamma$ cross section using BFKL dynamics
and found a characteristic rise with decreasing $x$, which
becomes steeper the more forward are the detected photons.

Our study should be regarded as an exploratory investigation of
the potential usefulness of the process.  The DIS + $\gamma$
rates which we present in Figs.\ 6 and 7 correspond to photons
produced (i) at more
than $5^\circ$ to the proton beam direction $(\theta_{\gamma p} >
\overline{\theta}_0 = 5^\circ)$ in the HERA frame, (ii) in the
centre of an isolation cone of half-angle $3^\circ$
($\theta_{\gamma q^\prime} > \theta_0 = 3^\circ$) in the HERA
frame, (iii) in the proton hemisphere in the $\gamma^* p$
c.m.\ frame (to avoid contamination with photons radiated from
the quark box), and (iv) with $k_{\gamma T}^2 > 5$ GeV$^2$.  Our
work is at the parton level and so in
practice the isolation criteria will need further study.  In
particular simulations of the fragmentation of the outgoing
($q^\prime$) jet should be performed so as to be able to choose
the optimum isolation criteria for the photon.  Recall that
isolation is required to suppress (background) photons from
$\pi^0$ decays.

There is a second reason why it is important to know the rate of
DIS + forward photon events.  We have stressed the experimental
difficulty of obtaining a clean sample of DIS events containing a
measured forward jet close to the proton remnants.  An
alternative possibility to expose $\log (1/x)$ resummation
effects is to use the improved knowledge of the fragmentation
functions to identify the forward jet through the measurement of
a single energetic decay product.  As it happens the $\pi^0$
is the hadron which can be identified in the most forward
direction in the detectors at HERA.  For this alternative signal
of small $x$ dynamics, our process is then the background. \\

\bigskip
\noindent {\large \bf Acknowledgements}

We thank Albert De Roeck, Genya Levin and Peter Sutton for
valuable help and discussions.  J.K.\ thanks the Department of
Physics and Grey College of the University of Durham for their
warm hospitality.  This work has been supported in part by Polish
KBN Grant No.\ 2 P03B 231 08 and the EU under Contracts Nos.\
CHRX-CT92-0004 and CHRX-CT93-0357.  SCL thanks the UK Engineering
and Physical Sciences Research Council for a Studentship. \\

\newpage
\section*{Figure Captions}
\begin{itemize}
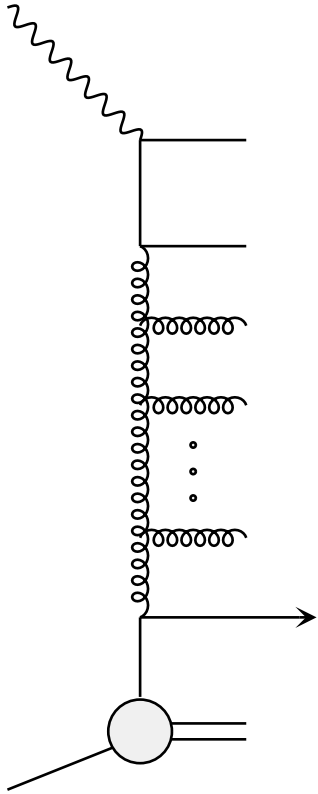
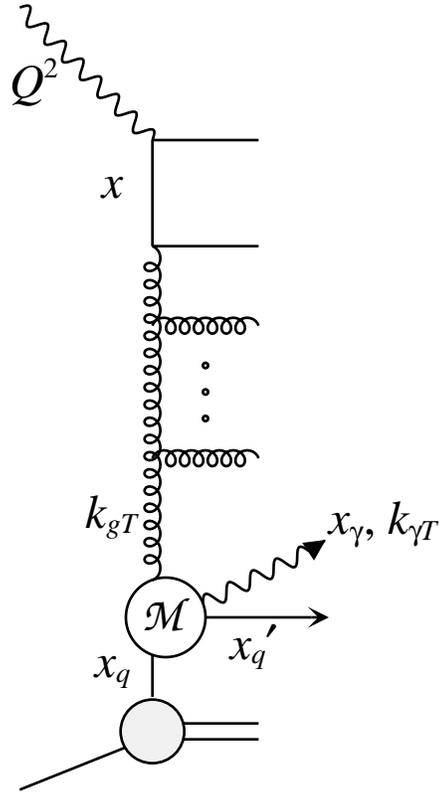
\item[Fig.\ 1] Diagrammatic representation of (a) a
deep-inelastic + forward jet event, and (b) a deep-inelastic $(x,
Q^2)$ + forward identified photon $(x_\gamma, k_{\gamma T})$
event.

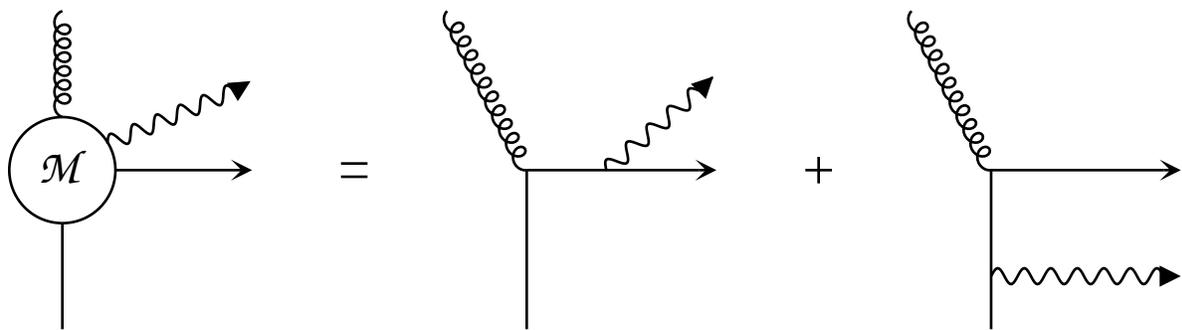
\item[Fig.\ 2] The Feynman diagrams describing the $gq
\rightarrow \gamma q$ subprocess embodied in the DIS + $\gamma$
diagram shown in Fig.\ 1(b).

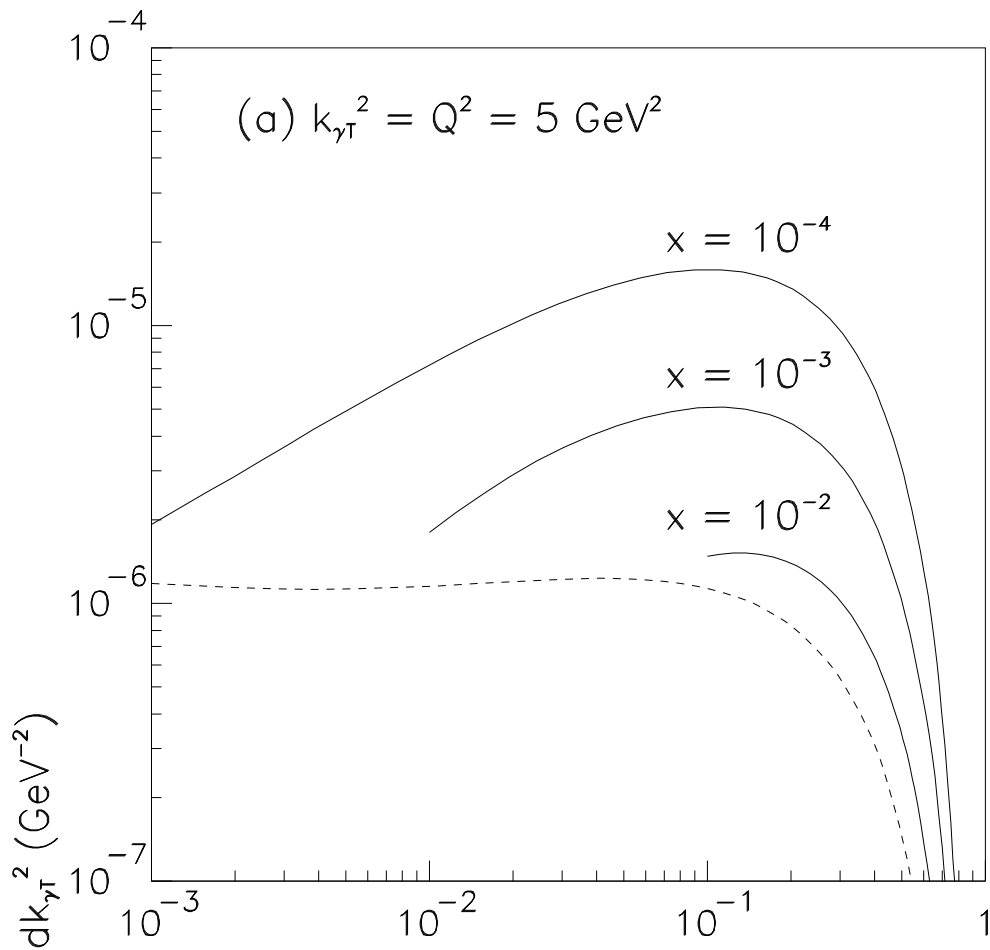
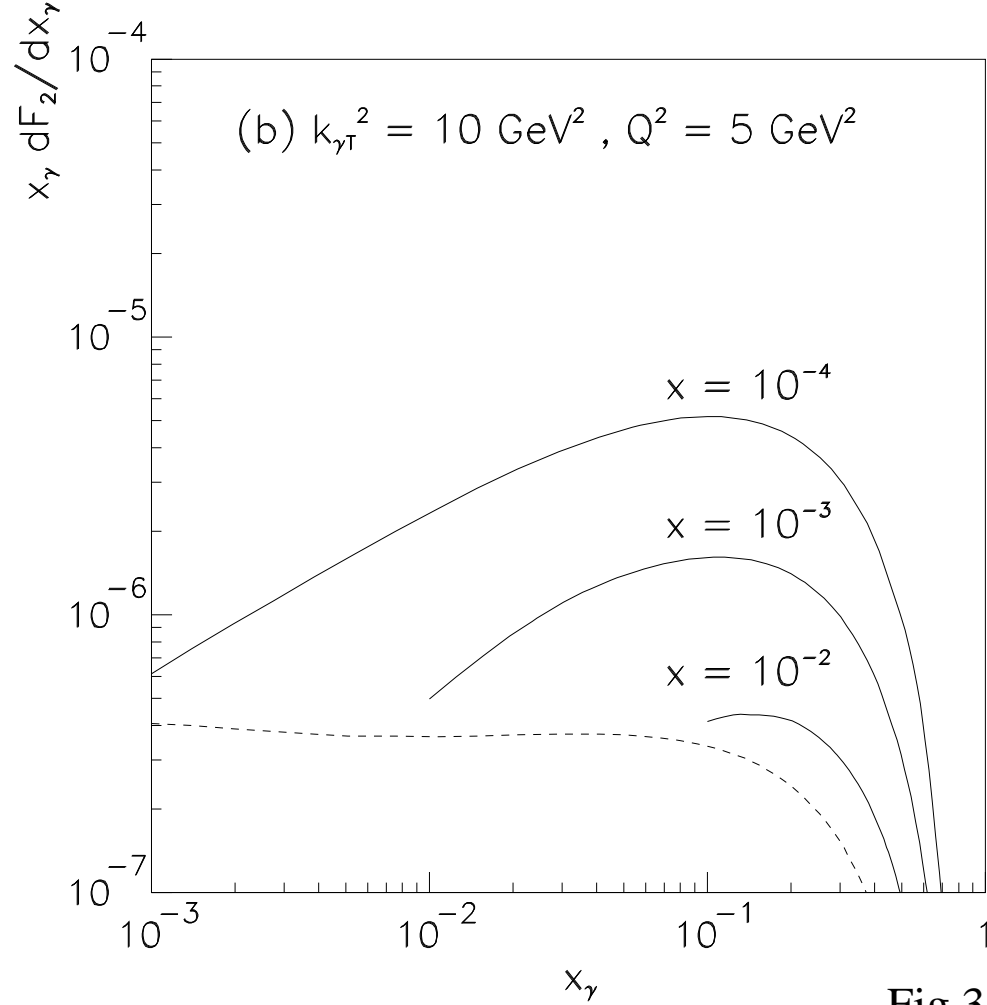
\item[Fig.\ 3] The $F_2$ differential structure function,
(\ref{eq:a4}), for deep-inelastic $(x, Q^2)$ events accompanied
by a measured forward photon $(x_\gamma, k_{\gamma T}^2)$ as a
function of $x_\gamma$ for different values of $x, x = 10^{-4},
10^{-3}$ and $10^{-2}$, and for $Q^2 = 5$ GeV$^2$.  The
continuous curves correspond to inputting into (\ref{eq:a4}) the
solution $\Phi_2$ of the BFKL equation (\ref{eq:a15}), whereas
the dashed curve is calculated using for $\Phi_2$ simply the
driving term $\Phi_2^{(0)}$ of (\ref{eq:a16}).  Here $|{\cal
M}|^2$ is regulated by requiring $\hat{s} > 1$ GeV$^2$.  Plots
(a) and (b) correspond to photons with $k_{\gamma T}^2 = 5$ and
10 GeV$^2$ respectively. 

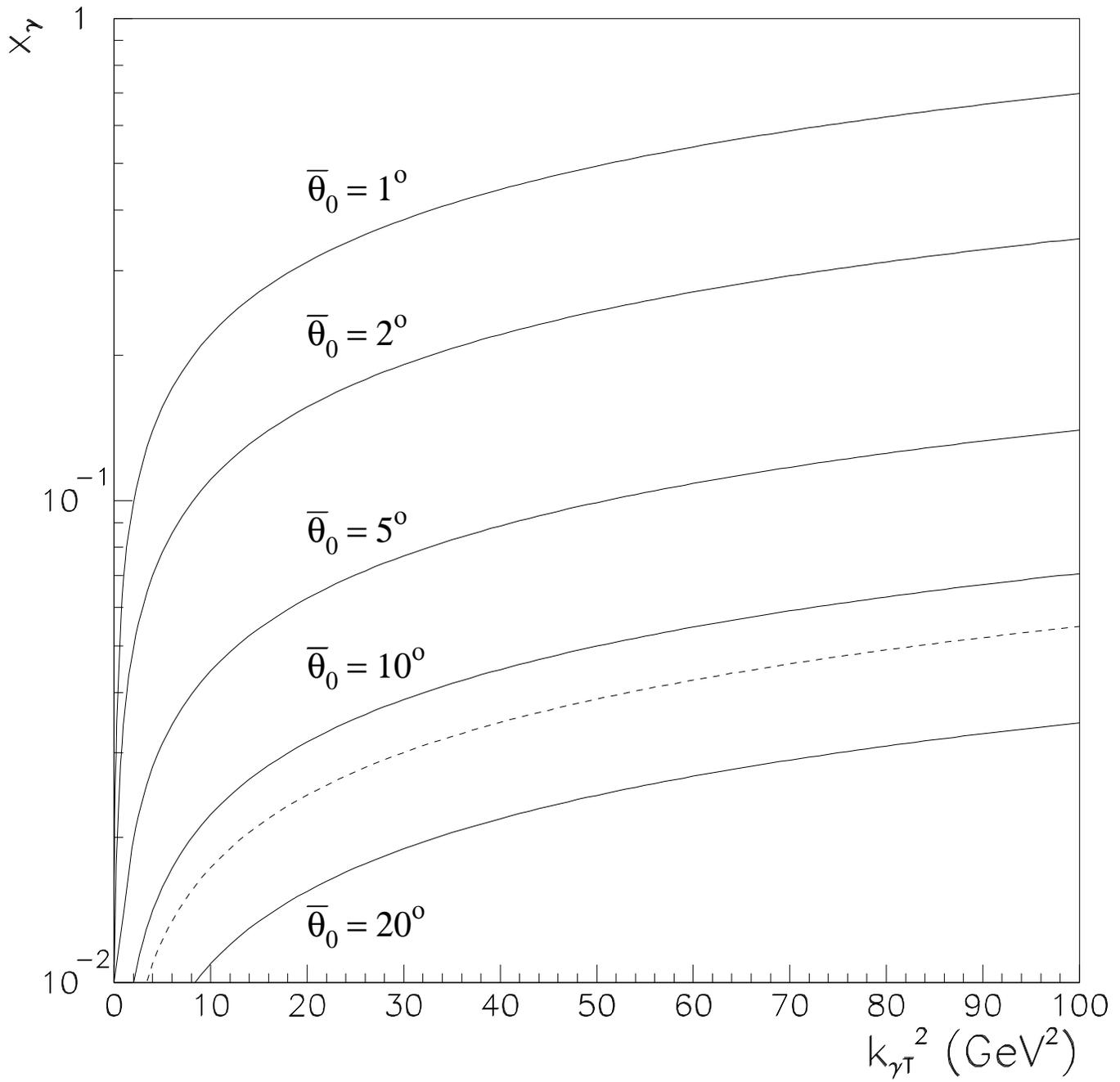
\item[Fig.\ 4] The curves give the {\it upper} boundary of the
allowed
regions of the photon kinematic variables $(x_\gamma, k_{\gamma
T}^2)$ for deep-inelastic + photon events with $x = 6 \times
10^{-4}$ and
$Q^2 = 20$ GeV$^2$ for various choices of the acceptance angle
$\overline{\theta}_0$ of (\ref{eq:a20}).  The photon angle
$\overline{\theta}_0$ to the proton direction in the HERA $(30
\times 820$ GeV) laboratory frame is not uniquely specified by
$(x, Q^2; x_\gamma, k_{\gamma T}^2)$.  Varying the remaining
azimuthal angle transforms the line of constant $\theta_{\gamma
p}$ into narrow bands in the $x_\gamma, k_{\gamma T}^2$ plane. 
The continuous lines that are shown are obtained by averaging
over the
azimuthal degree of freedom.  The lines are insensitive to
reasonable variations of $x$ and $Q^2$.  Also shown (by a dashed
line) is the {\it lower} boundary given by the hemisphere cut
(\ref{eq:a19}) for $x = 6 \times 10^{-4}$ and $Q^2 = 20$ GeV$^2$.

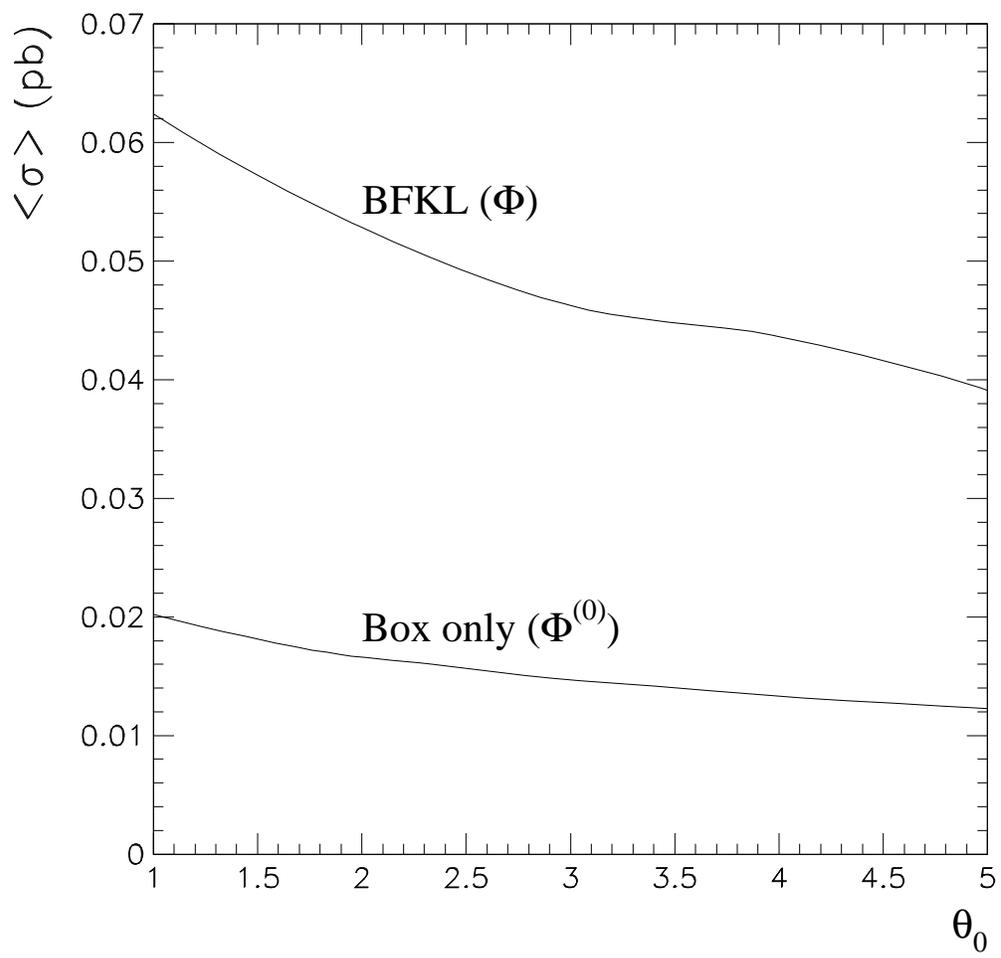
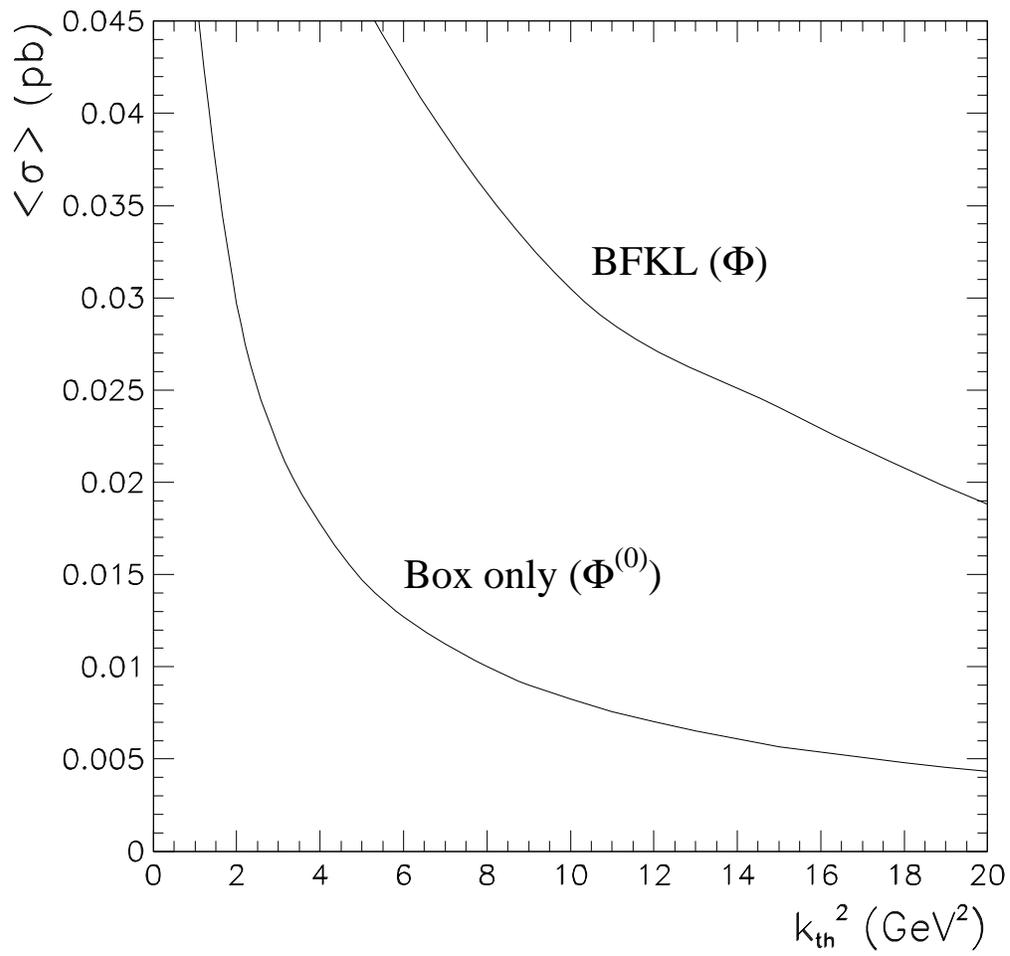
\item[Fig.\ 5] The dependence of the DIS + $\gamma$ cross section
integrated over the photon variables to variation of (a) the
angle $\theta_0$ defining the isolation cone around the photon
$(\theta_{\gamma j} > \theta_0)$ and (b) the threshold for photon
detection $(k_{\gamma T}^2 > k_{th}^2)$.  The results are for the
$x, Q^2$ bin defined by $6 \times 10^{-4} < x < 8 \times
10^{-4}$ and $20 < Q^2 < 30$ GeV$^2$.  We impose the hemisphere
cut (\ref{eq:b19}) and take $\overline{\theta}_0 = 5^\circ$ in
(\ref{eq:a20}).  (The lack of smoothness of the curves simply
reflects the errors on the six-fold numerical integration).

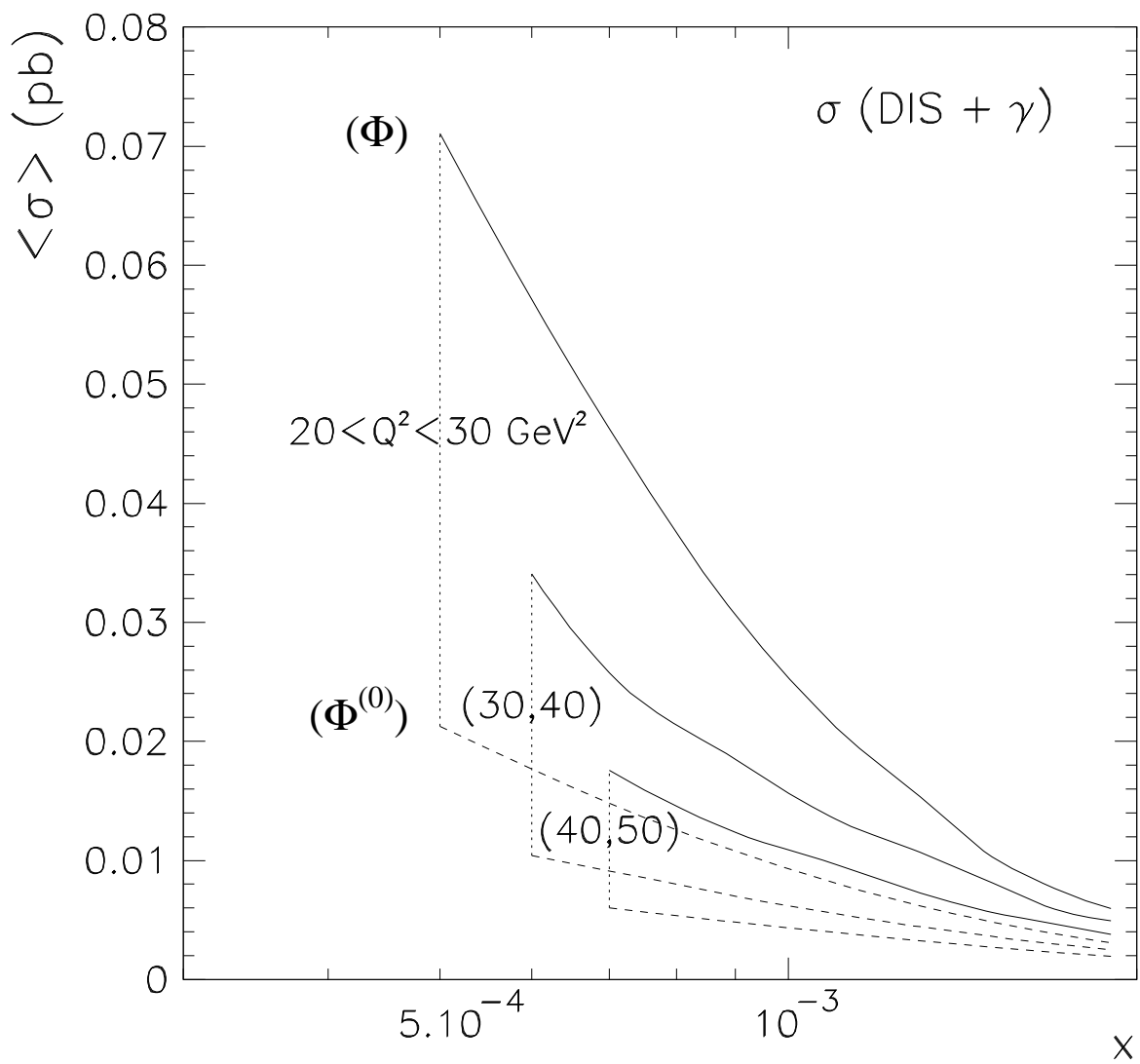
\item[Fig.\ 6] The cross section, $\langle \sigma \rangle$ in
pb, for deep inelastic + photon events integrated over $\Delta
x = 2 \times 10^{-4}$, $\Delta Q^2 = 10$ GeV$^2$ bins which are
accessible at HERA, and integrated over the region
$\theta_{\gamma p} > 5^\circ, k_{\gamma T}^2 > 5$ GeV$^2$, but
subject to (\ref{eq:b19}) and an isolation cut of $\theta_0 =
3^\circ$.  The $x$ dependence is shown for three different
$\Delta Q^2$ bins, namely (20,30), (30,40) and (40,50) GeV$^2$. 
The $\langle \sigma \rangle$ values are plotted at the central
$x$ value in each $\Delta x$ bin and joined by straight lines. 
The continuous curves show $\langle \sigma \rangle$ calculated
with $\Phi_i$ determined from the BFKL equation, whereas the 
dashed
curves are obtained just from the driving terms $\Phi_i^{(0)}$,
i.e.\ from the quark box.  For clarity a vertical line links the
pair of curves belonging to the same $\Delta Q^2$ bin.

\newpage
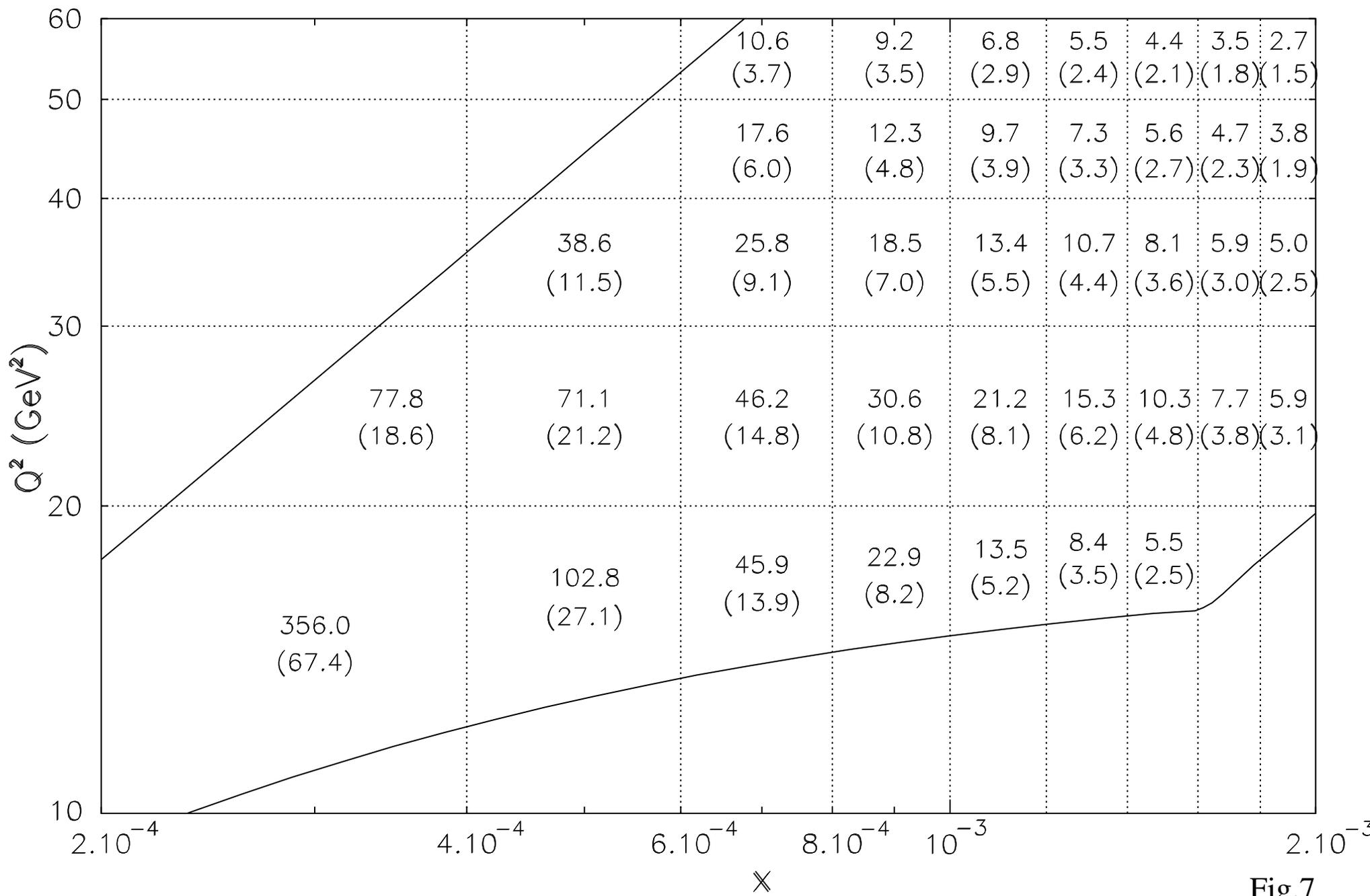
\item[Fig.\ 7] The cross section $\langle \sigma \rangle$ in fb
for deep inelastic + photon events in various $(\Delta x, \Delta
Q^2)$ bins which are accessible at HERA, and integrated over the
region $\theta_{\gamma p} > 5^\circ, k_{\gamma T}^2 > 5$ GeV$^2$,
but subject to (\ref{eq:b19}) and an isolation cut of $\theta_0 =
3^\circ$.  The number in brackets is the cross section calculated
with just the quark box approximation $(\Phi_i = \Phi_i^{(0)})$.
The difference between the two numbers is therefore the 
enhancement due to the BFKL soft gluon resummation.
\end{itemize} 
\end{document}